\documentclass[pre,preprint,showpacs,superscriptaddress]{revtex4}
\begin{document}
\bibliographystyle{apsrev}
\title{
Theoretical Analysis and Simulations of the Generalized Lotka-Volterra Model
}
\author{
Ofer Malcai}
\affiliation{Racah Institute of Physics,
The Hebrew University,
Jerusalem 91904,Israel}
\author{
Ofer Biham}
\affiliation{Racah Institute of Physics,
The Hebrew University,
Jerusalem 91904,Israel}
\author{
Peter Richmond}
\affiliation{
Department of Physics, Trinity College, Dublin, Ireland}
\author{
Sorin Solomon}
\affiliation{Racah Institute of Physics,
The Hebrew University,
Jerusalem 91904,Israel}

\begin{abstract}
The dynamics of generalized
Lotka-Volterra systems 
is studied by theoretical techniques and computer simulations.
These systems  
describe the time evolution of the wealth distribution of individuals 
in a society, as well as of the market values of firms in the stock market.
The individual wealths or market values are given by a set of time dependent variables
$w_i$, $i=1,...N$. 
The equations include a stochastic autocatalytic term (representing investments), 
a drift term (representing social security payments) and a time dependent 
saturation term (due to the finite size of the economy).  
The $w_i$'s turn out to exhibit a power-law distribution of the form
$P(w) \sim w^{-1-\alpha}$. 
It is shown analytically that 
the exponent $\alpha$ can be expressed as a function of one parameter, which is
the ratio between the constant drift component (social security) and the fluctuating
component (investments). 
This result provides a link between the 
lower and upper cutoffs of this distribution, namely between the
resources available to the poorest and those available to the richest in a given
society.
The value of $\alpha$ 
is found to be insensitive to variations in the saturation term,
that represent the expansion or contraction of the economy.  
The results are of much relevance to empirical studies that
show that the distribution of the individual wealth
in different countries during different periods in the 20th century 
has followed a power-law distribution with $1 < \alpha < 2$.
\end{abstract}


\maketitle

\newpage

\section{Introduction}

In recent years there has been considerable interest in the 
collection and analysis of
large volumes of economic data. 
Such data includes the
distributions of the income and wealth of individuals
\cite{Pareto1897,Mandelbrot1961,Mandelbrot1951,Mandelbrot1963,Atkinson1978,Levy1997,Takayasu1997,Jogi1998}, 
the market values of publicly traded companies as well as their short
and long term fluctuations
\cite{Mantegna1995,Gopikrishnan1998,Plerou1999,Gopikrishnan1999}.
A common observation is that distributions of economic data
exhibit a power-law 
behavior of the form

\begin{equation}
P(w) \sim w^{-1-\alpha}.
\label{eq:pareto}
\end{equation} 

\noindent
where the variable $w$ represents the wealth of an individual 
or market value of a company and $\alpha$ is the exponent that
provides the best fit to the empirical data.
Empirical studies show that the distribution the wealth of individuals 
in different countries follows the power-law behavior described by
Eq.~(\ref{eq:pareto}), 
with
$1 < \alpha < 2$
\cite{Pareto1897,Mandelbrot1951,Mandelbrot1961,Mandelbrot1963,Atkinson1978,Levy1997}.
These results stimulated theoretical studies in attempt to construct models
that reproduce the power-law behavior and predict the value of $\alpha$
\cite{Levy1995,Levy1996b,Solomon1996,Sornette1997,Zanette1997,Biham1998,Malcai1999}.

In this paper we study a stochastic dynamical model, based on the
Lotka-Volterra system
that gives rise to the power-law distribution of
Eq.~(\ref{eq:pareto}). 
The model consists of coupled 
dynamic equations which describe the discrete time evolution
of the basic system components 
$w_i$, $i=1,\dots,N$. 
The structure of these equations resembles the
logistic map and they are coupled through the average value
$\bar w(t)$.
The dynamics includes autocatalysis both at the individual level
and at the community level as well as a saturation term.
To model the non-stationary conditions
we introduce a time
dependent parameter into the saturation term in each of these equations.
We find that the system components spontaneously evolve into a 
power-law distribution 
of the form of Eq.~(\ref{eq:pareto}),
even in the presence of non-stationary external conditions.
Furthermore,
it is shown analytically that the exponent $\alpha$ 
depends only on the ratio of the constant drift component (social security) 
and the fluctuating component (investments).
It is found to be insensitive to variations in the saturation term
that describes the level of economic activity and varies
between periods of prosperity and depression.

The paper is organized as follows. 
In Sec. II we present the generalized Lotka-Volterra
model under non-stationary conditions. 
Analytical results and predictions are presented in Sec. III 
and compared with the results of numerical simulations in Sec. IV. 
A summary is presented in Sec. V. 

\section{The Model}
\label{model}

The generalized Lotka-Volterra system 
\cite{Levy1995,Levy1996b,Solomon1996,Biham1998,Malcai1999,Biham2001}
describes the evolution in discrete time of
$N$ dynamic variables
$w_i$, $i=1,\dots,N$.
In ecological systems, $w_i$ represents the population size of the $i$th
specie, while in economic systems it may represent the wealth of an individual
investor or the market value of a publicly traded firm. 
At each time step $t$, 
an integer $i$ is chosen randomly in the range
$1 \leq i \leq N$, which is the index of the dynamic variable $w_i$
to be updated at that time step.
A random multiplicative factor
$\lambda(t)$ is then drawn from a given distribution
$\Pi(\lambda)$, which is independent of $i$ and $t$.
It will later be convenient to express this multiplicative
factor by

\begin{equation}
\lambda(t) = \langle \lambda \rangle + \eta(t)
\end{equation}

\noindent
where $\langle \lambda \rangle$ is the average value of 
$\Pi(\lambda)$
and

\begin{equation}
D = \langle \lambda^2 \rangle - \langle \lambda \rangle^2
\end{equation}

\noindent
is its standard deviation.
The system is then updated according to

\begin{eqnarray}
w_i (t+1) &=&   [1+\lambda (t)]  w_i (t) 
+  \sum_{j=1}^N a_{i,j} w_j (t)  
-  \sum_{j=1}^N a_{j,i} w_i (t) 
- \sum_{j=1}^N c_{i,j} w_i (t) w_j(t)  \nonumber \\
w_j (t+1) &=& w_j (t), \ \ \ \ \ \ j=1,\dots,N; \ j \ne i. 
\label{eq:gediloij}
\end{eqnarray}

\noindent
where 
$a_{i,j}$
and
$c_{i,j}$
are constants.
This is an asynchronous update mechanism.
The first term on the right hand side of 
Eq.~(\ref{eq:gediloij}), 
describes the effect of stochastic auto-catalysis at the individual level. 
In an ecological system this term represents variations in the population
of a given specie, including births and deaths that 
may be affected by external conditions but
are not affected by
the interaction with other species.
In a stock-market system 
it represents the increase (or decrease) by a random factor 
$\lambda(t)$ 
of the capital of the investor 
$i$ between time $t$ and $t+1$.
The second and third terms in 
Eq.~(\ref{eq:gediloij}), 
describe the interaction between different dynamical variables. 
In an ecological system, the second term represents the dependence of population $i$
on the availability of food, in the form of population $j$. The third
term represents the fact that population $i$ itself may be the food of some
other species.
In an economic system the second and third terms represent trade between investors or 
firms $i$ and $j$, such as buying and selling, respectively.
The fourth term in 
Eq.~(\ref{eq:gediloij}), 
describes saturation effects due to the competition for limited resources. 
In an ecological model, this term implies that large 
populations tend to exhaust the available resources on which they depend. 
The saturation parameters $c_{i,j}$ are large for 
populations $i$ and $j$ that consume the same type of food.
In an economic system this term has to do with the 
saturation due to the finite size of the economy.

To simplify the analysis we will consider in this paper a simple case in which
the $w_i$'s interact in a uniform fashion with each other.
This case is obtained by choosing
$a_{i,j}=a/N$ 
and 
$c_{i,j}=c/N$.
With this choice 
Eq.~(\ref{eq:gediloij})
will be reduced to

\begin{eqnarray}
w_i (t+1) &=&   [1+\lambda (t)] w_i (t) +   a \bar w (t) - c w_i (t) \bar w(t)  \nonumber \\
w_j (t+1) &=& w_j (t), \ \ \ \ \ \ j=1,\dots,N; \ j \ne i. 
\label{eq:gedilo}
\end{eqnarray}

\noindent
where

\begin{equation}
\bar w(t) = {1 \over N} \sum_{i=1}^N w_i(t).
\label{eq:wbar}
\end{equation}

\noindent
is the average value of the dynamical variables at time t.
Here the random term $\lambda(t)$ was shifted to $\lambda(t)-a$ 
but its distribution around the average (and thus the value of the
standard deviation $D$) remain unchanged.
The second term in 
Eq.~(\ref{eq:gedilo}), 
may now describe the effect of  auto-catalysis at the community level. 
In an economic model, this term can be related to the social 
security policy or to general publicly funded services which every individual receives. 
It prevents an individual $w_i$ from falling below a certain fraction of the
average $\bar w$.
The third term in 
Eq.~(\ref{eq:gedilo}), 
describes  saturation or the competition for limited resources. 
It has the effect of limiting the growth to values sustainable for the
current conditions and resources.
Within the ecological context, here the interactions between populations
are uniform, describing the case in which all of them
consume the same type of food. 
We refer to 
Eq.~(\ref{eq:gedilo}) 
as the generalized Lotka-Volterra system
because when averaged over $i$
and over $\lambda(t)$,
this system tends to 
approach a
Lotka-Volterra-like 
equation
\cite{Lotka1925,Volterra1926} 

\begin{equation}
w (t+1) = ( { 1+\langle \lambda \rangle } +a ) w (t) - c w^2 (t).
\label{eq:lotkavolterra}
\end{equation}

\noindent
where $w(t) \equiv \bar w(t)$.
Computer simulations show that after some equilibration time 
the system described by
Eq.~(\ref{eq:gedilo})
approaches steady-state conditions.
Even at steady state,
$\bar w$ exhibits fluctuations.
However, its average over long time scales approaches a constant value
given by 

\begin{equation}
\langle \bar w \rangle_t  = ( \langle  \lambda \rangle +a) / c.
\label{eq:w_timeaveraged}
\end{equation}

In previous studies of the system described by
Eq.~(\ref{eq:gedilo}) the
parameters $a$ and $c$ 
were considered as constants, corresponding to steady
conditions of the market.
In fact, the typical dynamics of microscopic market models
\cite{Levy1994,Levy1995,Solomon1996,Levy1997} 
is generically {\it not} in a steady state.
The effect of varying market conditions can be
studied by considering the parameters 
$a$ and $c$ and the distribution  
$\Pi (\lambda)$ 
as slowly varying functions of time.
We will show below that  
systems described by 
Eq.~(\ref{eq:gedilo}) 
lead, under very general conditions, 
to a power-law distribution of the $w_i$'s 
of the form of 
Eq.~(\ref{eq:pareto}).
Moreover, it will be shown that the exponent $\alpha$
is insensitive to variations in the
parameter $c$, namely it depends only on $a$ and $D$. 
In order to examine the effect of variations in the economic conditions
we will now introduce an explicit time dependence into the third term,
as well as a more general dependence on the $w_j$'s.
The dynamic equation will now take the form

\begin{eqnarray}
w_i(t+1) &=& [1+\lambda (t)] w_i(t) + a \bar w(t) - C(w_1,\dots,w_n,t) w_i(t) \nonumber \\
w_j (t+1) &=& w_j (t), \ \ \ \ \ \ j=1,\dots,N; \ j \ne i, 
\label{eq:gedilocontgen} 
\end{eqnarray}

\noindent
where $C(w_1,\dots,w_n,t)$ is a general function of the $w_j$'s
that includes an explicit time dependence.

\section{Theoretical Analysis}
\label{theoretical}

In order to study the dynamics of the generalized Lotka-Volterra model, it will 
be convenient to 
denote the change of $w_i$ in a single time step by
$\Delta w_i(t) = w_i(t+1) - w_i(t)$.
We introduce a set of normalized variables

\begin{equation}
x_i = {w_i \over \bar w}, \ \ \ i=1,\dots,N.
\label{eq:x_i}
\end{equation}

\noindent
The change 
$\Delta x_i(t) = x_i(t+1) - x_i(t)$
in a single time step
is given by

\begin{equation}
\Delta x_i \cong  {1 \over \bar w}  \Delta w_i - {w_i \over {{ \bar w}^2}} 
\Delta {\bar w}
\label{eq:dx_dt}
\end{equation}

\noindent
up to first order in powers of the $\Delta w_i$'s.
Considering the 
time dependence of the average $\bar w$ 
one should remember that at any time step $t$ of 
Eq.~(\ref{eq:gedilocontgen})
only one of the $w_i$'s is chosen randomly and updated. 
Moreover, there is no correlation between the chosen
$w_i$ and $\lambda(t)$.
Thus, the time evolution of $\bar w$ should be considered
on a longer time scale of the order of $N$ moves. 
However, for simplicity we evaluate $\Delta {\bar w}$ by 
averaging 
Eq.~(\ref{eq:gedilocontgen})
for $\Delta w_i$
over 
$i=1,\dots,N$ 
at a given time $t$. 
We make an independent
random choice of $\lambda(t)$ for each $i$, 
which we denote by
$\lambda_i(t) = \langle \lambda \rangle + \eta_i(t)$.
The time dependence of $\bar w$ is given by

\begin{equation}
\Delta {\bar w} =  {{1} \over {N}} \sum_{j=1}^N { \eta}_{j} (t) w_j (t) + \left[ \left<
 \lambda \right> + a \right] \bar w(t) - C(w_1,\dots,w_n,t) \bar w(t).    
\label{eq:dwbar_dt}
\end{equation}

\noindent
The dynamics of the $x_i$'s is thus given by 

\begin{equation}
\Delta x_i = x_i(t) \left[ \eta (t) - a - 
{{1} \over {N}} \sum_{j} { \eta}_{j} (t) x_j (t) \right] + a. 
\label{eq:dx}
\end{equation}

\noindent
Consider the sum

\begin{equation}
r(N) = \sum_{j=1}^N { \eta}_{j} (t) x_j (t).
\label{eq:levyflight}
\end{equation}

\noindent
If the $x_j$'s exhibit a distribution 
of the form

\begin{equation}
P(x) \sim x^{-1-\alpha}, 
\label{eq:powerx}
\end{equation}

\noindent
then the second moment of the distribution
of $r(N)$ satisfies
\cite{Shlesinger1993,Shlesinger1994}

\begin{equation}
\langle r^2(N) \rangle ^{1/2} = 
\left\{
\begin{array}{lll}
N^{1/2},                     & 2 < \alpha             \\
N^{(3-\alpha)/2},            & 1 < \alpha < 2         \\ 
N,                           & 0 < \alpha < 1.          
\end{array}
\right.
\label{eq:2ndMom}
\end{equation}

\noindent
In the first case, the distribution of the $x_j$'s exhibits a finite
second moment and $r(N)/N \rightarrow 0$ in the limit $N \rightarrow \infty$.
In the second case the second moment of $P(x)$ diverges and
$r(N)$ follows a L\'evy distribution.

In both cases, namely for
$\alpha > 1$, we obtain that
in the (thermodynamic) limit
$N \rightarrow \infty$:

\begin{equation}
{1 \over {N}} x_i \sum_{j} { \eta}_{j} (t) x_j (t) \rightarrow 0.
\label{eq:sumgotozero}
\end{equation}

\noindent
Thus, 
under the assumption that $P(x)$ follows 
Eq.~(\ref{eq:powerx})
with $\alpha>1$, 
we obtain
to a good approximation that 
for large values of $N$ 

\begin{equation}
\Delta x_i = \left[ \eta (t) - a \right] x_i + a, \ \ \ i=1,\dots,N. 
\label{eq:gedilo_mf}
\end{equation}

\noindent
We see that the dynamics of the normalized variable $x_i$ is reduced to a set of 
identical decoupled linear Langevin equations,
which do not depend on the function 
$C(w_1,\dots,w_n, t)$ 
or on the 
mean value 
$\langle \lambda \rangle$ 
of the multiplicative noise.
These equations can be cast into 
a general framework of multiplicative processes
of the form  

\begin{equation}
\Delta x(t) = \eta (t) G(x (t)) + F(x (t)).
\label{eq:langevin}
\end{equation}

\noindent
Eq.~(\ref{eq:gedilo_mf})
can then be recovered 
by taking  $F(x_i) = a(1-x_i)$ and $G(x_i)=x_i$. 
By using a suitable change of variables to 
$y=y(x)$
that satisfies

\begin{equation}
{d y \over d x}  = {1 \over G(x)} 
\label{eq:xtoy}
\end{equation}

\noindent
one can reduce
Eq.~(\ref{eq:langevin})
to a Langevin equation in which the term $\eta(t)$ appears as an additive 
noise, rather than a multiplicative noise such as
$\eta (t) G(x (t))$
\cite{Richmond2001}.
The time evolution of $y(t)$ 
is obtained from 
Eq.~(\ref{eq:langevin}) 
by using the chain differential rule up to second order 
in $\Delta x$
(and first order in $D$)

\begin{equation}
\Delta y \simeq {d y \over dx } \Delta x + {1 \over 2}{d^2 y \over dx^2 } {\Delta x}^2.  
\label{eq:chain1}
\end{equation} 

\noindent
Inserting $\Delta x(t)$ from 
Eq.~(\ref{eq:langevin})
and using the change of variables
described in 
Eq.~(\ref{eq:xtoy})
we obtain

\begin{equation}
\Delta y \simeq {1 \over G} (F+\eta G) + 
{1 \over 2} {d \over dx} ({1 \over G}) (F + \eta G)^2.
\label{eq:chain2}
\end{equation} 

\noindent
We now approximate the second order term by averaging over
the noise term $\eta$,
that satisfies  
$\langle \eta \rangle =0$
and
$\langle \eta^2 \rangle = D$.
We obtain

\begin{equation}
\Delta y \simeq  \eta + {F \over G} 
- {1 \over 2} {dG \over dx}  (D +  {F^2 \over G^2}).
\label{eq:langevinD}
\end{equation} 

\noindent
Assuming that $F^2/G^2 \ll D$, 
Eq.~(\ref{eq:langevinD})
is reduced to a discrete-time Langevin equation 

\begin{equation}
\Delta y \simeq  
\eta + J(y)
\label{eq:langevinJ}
\end{equation} 

\noindent
where the drift force $J(y)$ takes the form 

\begin{equation}
J(y) = {F \over G} - {D \over 2} {d G \over d x}.
\label{eq:J}
\end{equation}

\noindent
The Fokker-Planck equation corresponding to 
Eq.~(\ref{eq:langevinJ}) 
is
\cite{vanKampen1981}

\begin{equation}
{{\partial P(y,t)} \over {\partial t}} = 
- {{\partial} \over {\partial y}} {\left[J(y,t) P(y,t) \right]} 
+ {D \over 2}{{{\partial}^2 P(y,t)} \over {\partial y^2}}
\label{fokker-planck}
\end{equation} 

\noindent
where $P(y,t)$ is the probability distribution of $y$ at time $t$.
The solution of this equation under
the stationary condition 
${{\partial P(y,t)} / {\partial t}} = 0$ 
is

\begin{equation}
P(y) = \exp \left[ {2 \over D} \int^{y} J (y^{\prime}) dy^{\prime} \right].
\label{maxwell}
\end{equation}

\noindent
Thus, the distribution 
$P(x) = P(y) dy/dx$
of the original variables 
$x_i$ 
is

\begin{equation}
P(x) = {1 \over G^2(x)} 
\exp \left[ {2 \over D} \int^{x} 
{ F(x^{\prime}) \over G^2(x^{\prime})} dx^{\prime} \right].
\label{eq:x_dis}
\end{equation}

\noindent
By taking 
$F(x) = a(1-x)$ and $G(x) = x$ 
we obtain a power-law distribution of the form
\cite{Solomon2001}

\begin{equation}
P(x) = x^{-1 - \alpha} \exp \left[ {1-\alpha} \over x \right].
\label{eq:power-law}
\end{equation} 

\noindent
where

\begin{equation}
\alpha = 1 + {2a \over D}.
\label{eq:alpha}
\end{equation}

\noindent
The exponent $\alpha$ thus depends on a single parameter $a/D$, namely
on the ratio between the global drift coefficient $a$ and the
fluctuations measured by $D$. 

Another way to derive 
Eq.~(\ref{eq:alpha})
from dynamical models of the form
(\ref{eq:gedilo_mf})
was shown in Refs.
\cite{Takayasu1997,Sornette1997}.
It is based on the fact that, 
under steady-state conditions,
linear Langevin equations
of the form
(\ref{eq:gedilo_mf})
satisfy
\cite{Takayasu1997,Sornette1997,Kesten1973}

\begin{equation}
\langle (\eta -a +1)^{\alpha} \rangle =1.
\label{eq:Kesten}
\end{equation}

\noindent
Considering $\eta - a$ as a small parameter and expanding
Eq.~(\ref{eq:Kesten})
in a power series up to second order
we obtain

\begin{equation}
\alpha = 1 + {{2 a} \over {D + a^2}}.
\end{equation}

\noindent
Assuming that $a^2 \ll D$ we reproduce 
Eq.~(\ref{eq:alpha}).

\section{Numerical Simulations and Results}
\label{simulations}

To examine the theoretical predictions presented in Sec. III
we have performed computer simulations of the generalized Lotka-Volterra
system described by 
Eq.~(\ref{eq:gedilocontgen})
with different choices of
$C(w_1,\dots,w_n, t)$.
It was found that after some equilibration time 
the distribution $P(x)$
reaches a steady state, 
and exhibits a power-law behavior.
For
$C(w_1,\dots,w_n, t)=c \bar w$
[as in 
Eq.~(\ref{eq:gedilo})],
$\bar w(t)$
fluctuates around some average value,
given by 
Eq.~(\ref{eq:w_timeaveraged}).
For a general function
$C(w_1,\dots,w_n, t)$,
that exhibits an explicit time dependence,
$\bar w(t)$ continues to vary according to this
function and its temporal average does not reach a steady state.

To examine how robust the power-law distribution is 
under varying conditions we have simulated 
Eq.~(\ref{eq:gedilocontgen})
with

\begin{equation}
C(w_1,\dots,w_n, t) =  c_o \left(1 + \sin {2 \pi t \over T}\right) \sum_{j=1}^N w_j^2
\label{eq:Ctimedep}
\end{equation}

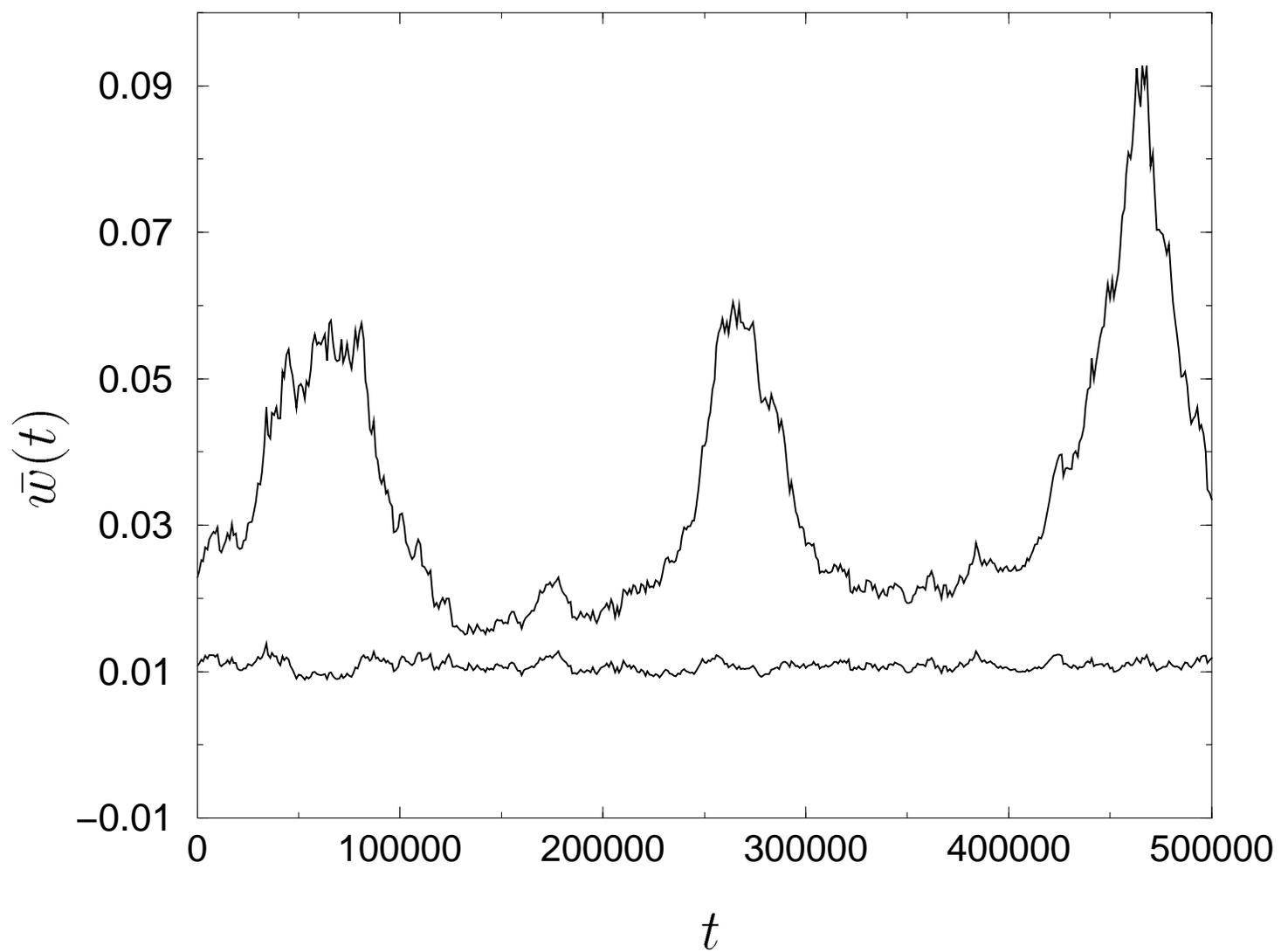
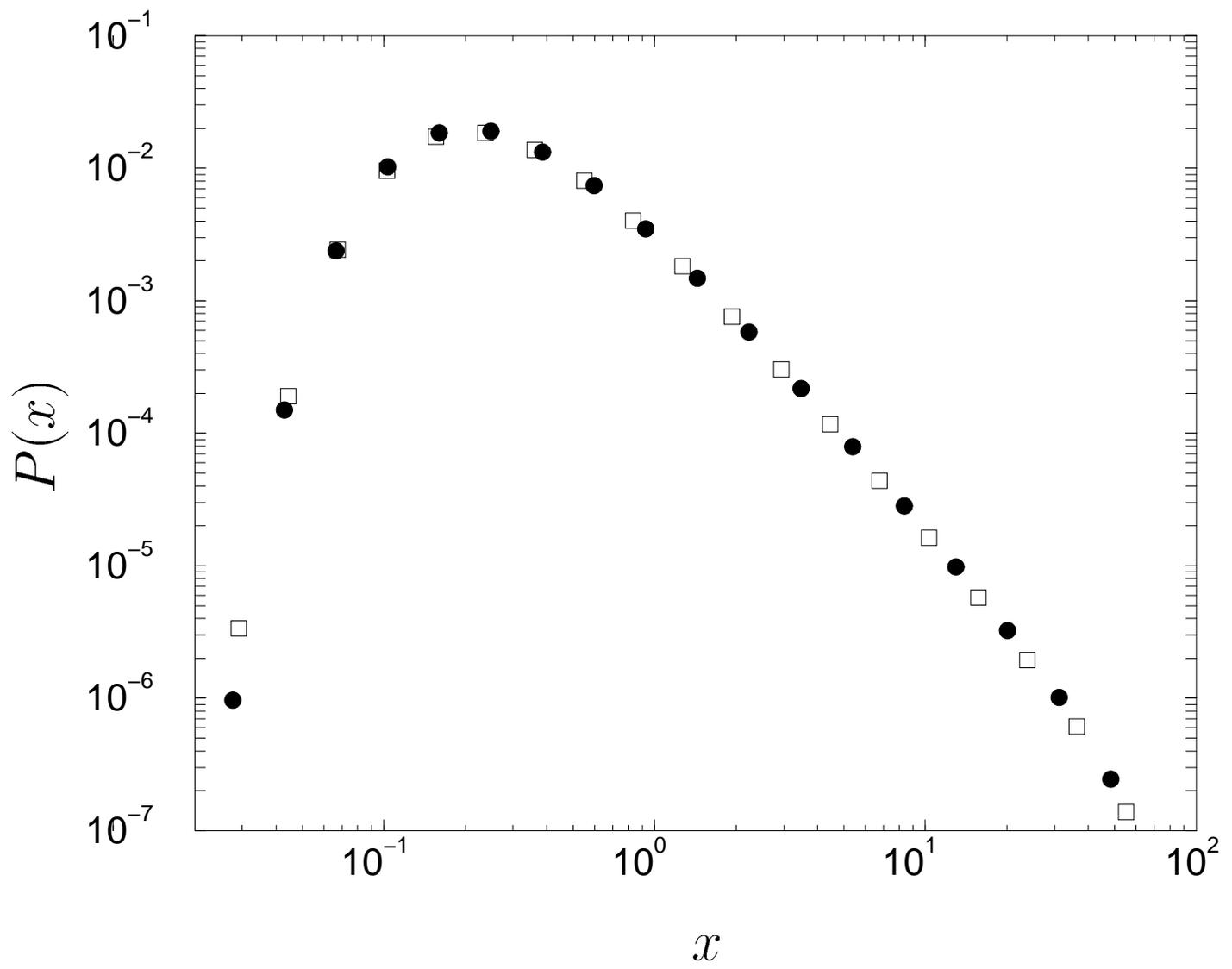
\begin{figure}
\caption{\label{fig:compare}
(a) The time dependence of the average wealth $\bar w$ for
the model of 
Eq.~(\ref{eq:gedilocontgen})
with
$C(w_1,\dots,w_n, t)$
given by 
Eq.~(\ref{eq:Ctimedep})
where
$c_0=0.001$
and
$T=2 \times 10^5$
(upper curve),
and with
$C(w_1,\dots,w_n, t) = c \bar w$
where $c=1$
(lower curve).
In the first case 
$\bar w$ 
oscillates, following the time dependence of
$C(w_1,\dots,w_n, t)$,
while in the second case it only
exhibits small fluctuations around a constant
value. 
In both cases
$N=100$, $a=0.00083$,
$D=0.0033$;
(b)
The distributions of the variables $x_i = w_i/ \bar w$
for the simulations shown in the lower curve (squares)
and the upper curve ($\bullet$) in (a).
The two distributions are found to be nearly identical, 
showing an approximate power-law behavior.
We thus observe that the exponent $\alpha$ 
is robust and insensitive to variations in 
$C(w_1,\dots,w_n, t)$. 
}
\end{figure}

\noindent
(for $c_0=0.001$, $T=2 \times 10^5$ and $\langle \lambda \rangle = 0.002$)
and compared the results with the case
$C(w_1,\dots,w_n, t)=c \bar w$
(for $c=1$ and $\langle \lambda \rangle=0.01$).
The time dependence of $\bar w$ in both cases is shown
in 
Fig.~\ref{fig:compare}(a). 
The distributions $P(x)$ obtained 
from the simulations
in these two cases 
are shown in 
Fig.~\ref{fig:compare}(b). 
The distributions are found to be nearly identical and exhibit a power-law behavior
characterized by the same exponent $\alpha$.
The exponent $\alpha$ is also found to
be independent of
$\langle \lambda \rangle$.
Note that the power-law behavior is maintained even for 
$C(w_1,\dots,w_n, t) \equiv 0$, 
where $\bar w(t)$
does not reach a steady state and diverges to infinity (for $\langle \lambda \rangle > 0$)
or collapses to $0$ (for $\langle \lambda \rangle < 0$)
\cite{Malcai1999}.
This can lead to changes by orders of magnitude in the total wealth or the 
population size without 
affecting the exponent $\alpha$.

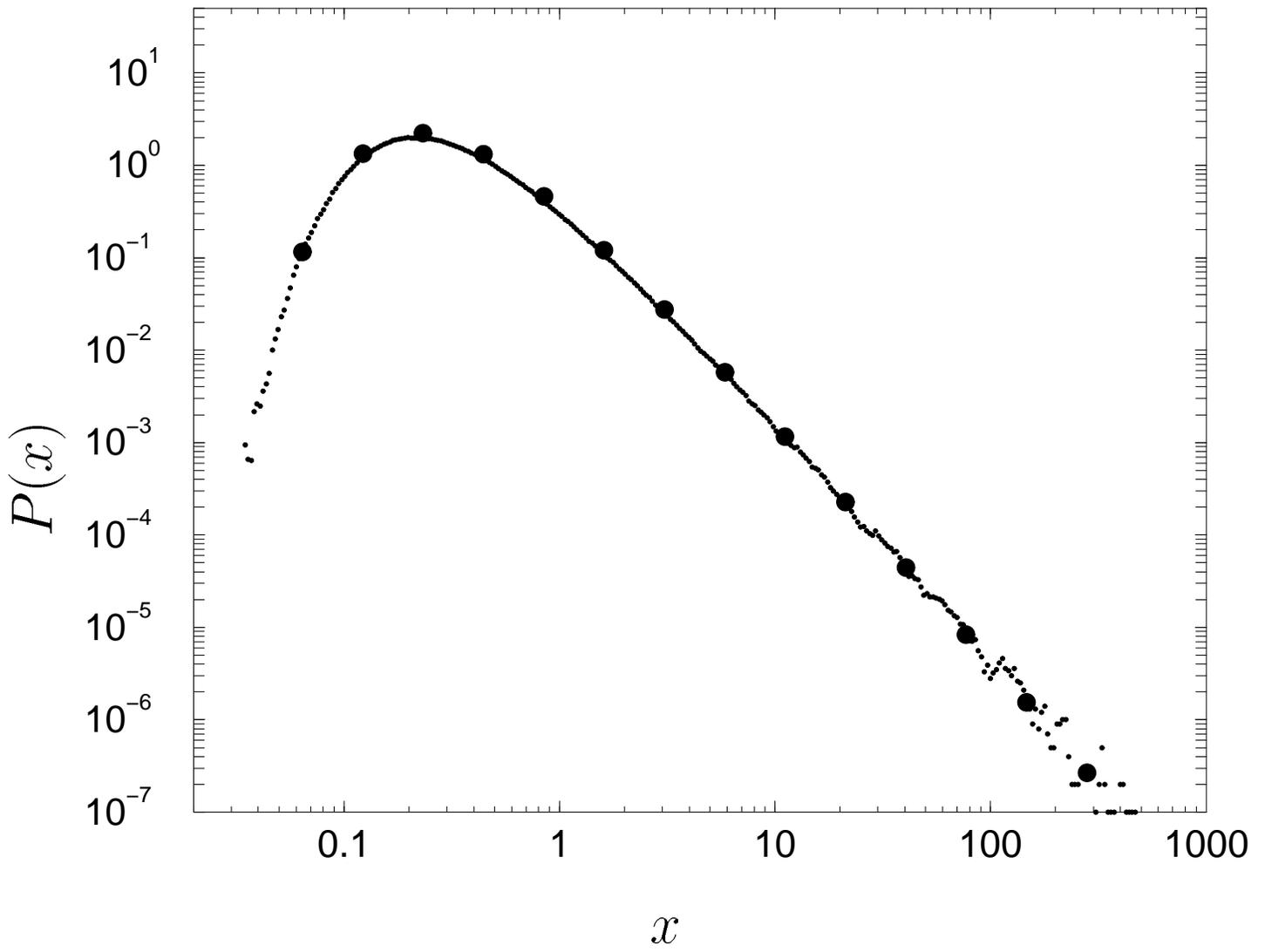
\begin{figure}
\caption{\label{fig:pareto_analytic}
Results of computer simulations (dots) and theoretical analysis based
on 
Eq.~(\ref{eq:power-law})
($\bullet$)
for 
the distribution of the variables 
$x_i = w_i/ \bar w$.
The parameters are
$N=1000$, $a=0.00023$, 
$D=0.00083$ 
and
$C(w_1,\dots,w_n,t) = 0.01 \bar w$. 
In both cases a power-law distribution is obtained with an excellent agreement
between the theoretical predictions and the simulation results.
}
\end{figure}

To examine the theoretical prediction for the
distribution, given by 
Eq.~(\ref{eq:power-law}),
and the exponent $\alpha$,
given by 
Eq.~(\ref{eq:alpha}),
we have compared these predictions to the results
of numerical simulations.
This comparison for the distribution $P(x)$
is shown in 
Fig.~\ref{fig:pareto_analytic}.
The simulations were done for
$N=1000$, $a=0.00023$, $C(w_1,\dots,w_n, t)=0.01 \bar w$, and $\lambda(t)$ 
uniformly distributed in the range
$0.0 \leq \lambda(t) \leq 0.1$ (namely, $D=0.00083$).
We found that the simulation results are in very good agreement with
the theoretical predictions. 
A power law distribution is found for a range
of almost three orders of magnitude, with
the exponent $\alpha=1.52$.
This is
close to the theoretical prediction of
$\alpha = 1 + 2a/D = 1.55$.
The distribution has a peak at 
$x_0 = (\alpha -1)/(\alpha+1)$,
that using 
Eq.~(\ref{eq:alpha})
can be expressed by
$x_0 = a / (a + D)$. 
Above $x_0$ the distribution $P(x)$
behaves like power-law while below it $P(x)$ decays exponentially.
This provides an effective lower cutoff for the range of $x$
in which a power-law behavior is observed.
This result can be compared 
to a somewhat simpler model studied 
earlier, in which the value of the lower cutoff $x_{min}$ is imposed as
a constraint
\cite{Malcai1999}.
In this model, using the sum rules for the probability and the total
wealth, it was found that
$x_{min}=1 - 1/\alpha$.
Using 
Eq.~(\ref{eq:alpha}) 
it can be expressed as
$x_{min} = 2 a /(2a + D)$. 
These predictions for the lower bounds in the two models 
satisfy
$x_0 < x_{min} < 2 x_0$,
namely, they are in good agreement
in light of the broad distribution of $x$.

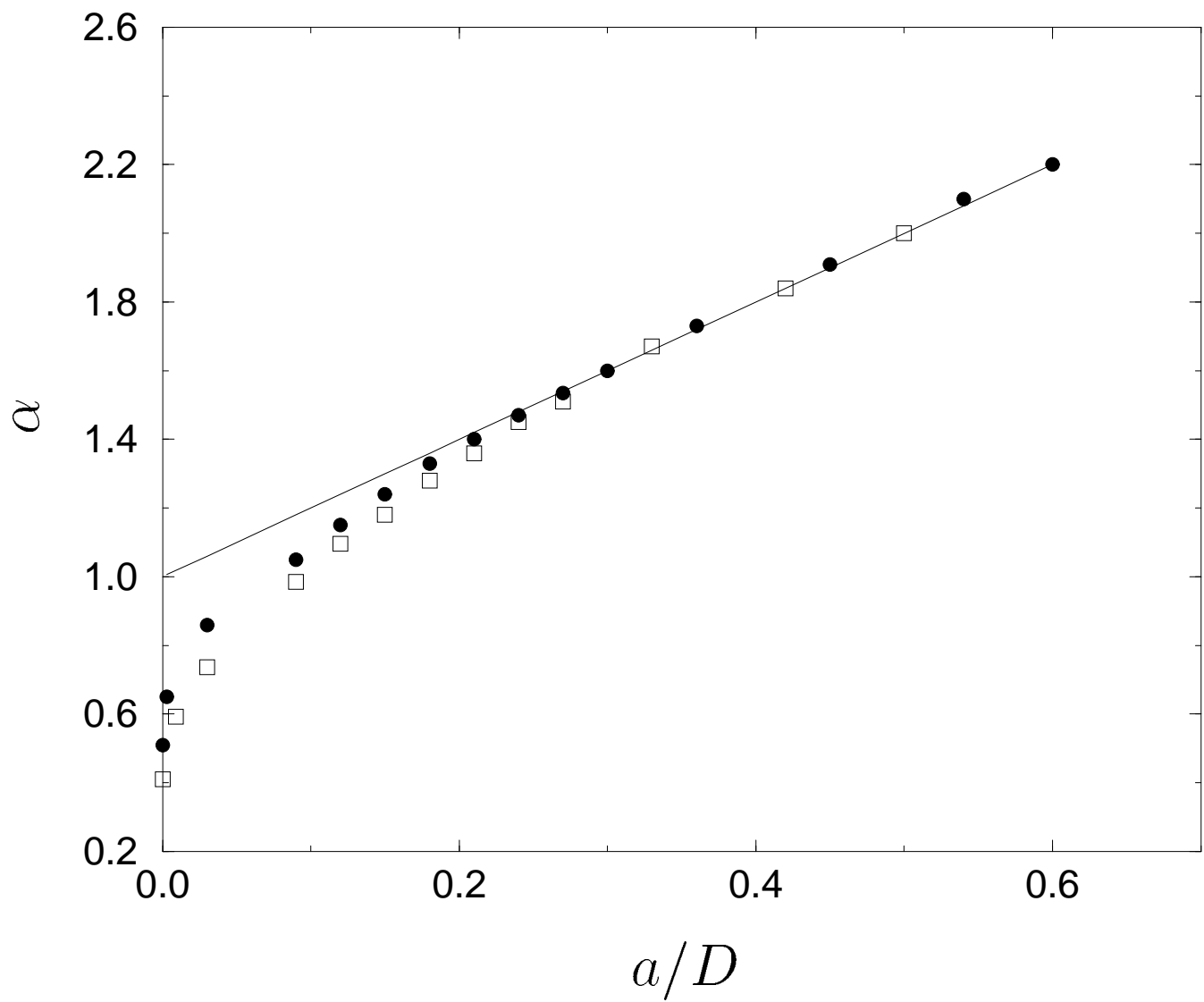
\begin{figure}
\caption{\label{fig:alpha}
Simulation results for
the exponent $\alpha$ of the power-law distribution of the 
variables $x_i = w_i/ \bar w$, $i=1,\dots,N$
as a function of the parameter $a/D$
for $N=1000$ and  $C(w_1,\dots,w_n,t) = 0.00001 \bar w$
($\bullet$).
The theoretical prediction of 
Eq.~(\ref{eq:alpha}) (line) is found to be in 
agreement with the numerical values
for $a/D > 0.2$. The agreement for small values of $a/D$ tends
to improve as $N$ increases.
}
\end{figure}

To examine the prediction given by 
Eq.~(\ref{eq:alpha})
for the exponent $\alpha$,
we present in 
Fig.~\ref{fig:alpha}
a comparison between this prediction and the numerical results
for $\alpha$ as a function of $a/D$.
The numerical results are presented 
for 
$N=1000$ ($\bullet$).
The prediction of 
Eq.~(\ref{eq:alpha}) 
(solid line), 
shows a good agreement with the numerical 
results for $a/D > 0.2$.
The numerical results for the range of small $a/D$
converge to the theoretical prediction as the value
of $N$ is increased.
As shown in Ref.
\cite{Malcai1999}
the infinite system limit, $N \rightarrow \infty$, 
and the vanishing coupling limit, $a/D \rightarrow 0$,
do not commute.
On the one hand,
for any finite $N$ and $a/D \rightarrow 0$ the exponent
$\alpha \rightarrow 0$. 
On the other hand, for any fixed positive value of $a/D$ (no matter how small)
and $N \rightarrow \infty$ the exponent
$\alpha \ge 1$.
The majority of empirical results 
in which $1<\alpha<2$,
indicate that the second case is
highly relevant
and that the theoretical predictions of 
Eqs.~(\ref{eq:power-law})
and
(\ref{eq:alpha})
broadly apply.
Thus,
for $N \rightarrow \infty$, both the exponent $\alpha$ of the power-law decay, 
and the lower bound $x_0$ depend only on a single parameter $a/D$.
In the economic context, this parameter represents the ratio between 
the fixed income of minimal wage jobs or social security payments and
the level of fluctuations of the speculative income/loss.

\section{Summary}

We have studied
the dynamics of stochastic 
Lotka-Volterra systems 
under non-stationary conditions using both analytical and numerical
techniques.
For this class of  models, we found that in order to obtain a 
power-law distribution, 
it is sufficient that relative returns of the agents 
are stochastically equivalent. The assumption that the distribution 
$\Pi(\lambda)$ of the multiplicative noise, is independent of $i$, means 
that there are no investors or strategies that can obtain 
'abnormal' returns. This can be related to to the 'efficient market hypothesis', 
which assumes that the market pricing mechanism is so efficient that it reaches 
the 'right price' before any of the agents can take systematic advantage.
Therefore, the presence of a power-law distribution may be a sign of 'market efficiency', 
by analogy with Boltzmann distributions in statistical mechanics systems, 
which characterize thermal equilibrium.
Here we have shown that the power-law distribution is
stable even under non-stationary economic conditions, 
that are represented by the time dependence of
the saturation term $C(w_1,\dots,w_n, t)$.
We found that even under such conditions the distribution of the
(normalized) dynamical variables $x_i$ follow a power-law distribution
with an exponent $\alpha$. 
An expression for $\alpha$ in terms of ratio of the parameters $a$ and $D$
was obtained
[Eq.~(\ref{eq:alpha})].
In the economic context,
the parameter
$a$ represents the minimal wage or social 
security payments, while $D$ represents the level of fluctuations
in speculative income/loss. 
These results provide the distribution of wealth in a society
in terms of the social security policy and the volatility of the stock 
market. They also provide a connection between the incomes/wealths of the
poorest and the richest sectors of the society as a function of a
single parameter.

\newpage

\end{document}